\documentclass[arXiv]{actapoly}

\usepackage[active]{srcltx}
\usepackage{textcomp}
\usepackage{amsfonts}
\usepackage{graphicx}
\usepackage{graphics}

\newcommand{\be}{\begin{equation}}
\newcommand{\ee}{\end{equation}}
\newcommand{\ba}{\begin{array}}
\newcommand{\ea}{\end{array}}
\newcommand{\bea}{\begin{eqnarray}}
\newcommand{\eea}{\end{eqnarray}}

\def\C{{\mathbb C}} 
\newcommand{\bi}{\begin{itemize}}
\newcommand{\ei}{\end{itemize}}

\newtheorem{theorem}{Theorem}

\newtheorem{example}{Example}

\begin{document}

\title[Laplace equations]{Laplace equations, conformal superintegrability and B\^ocher contractions}

\author[E. Kalnins]{Ernest Kalnins}{their}
\correspondingauthor[W. Miller]{Willard Miller, Jr}{my}{miller@ima.umn.edu}
\author[E. Subag]{Eyal Subag}{theirs}

\institution{their}{Department of Mathematics, University of Waikato, Hamilton, New Zealand}
\institution{my}{School of Mathematics, University of Minnesota, Minneapolis, Minnesota 55455, USA}
\institution{theirs}{Department of Mathematics, Pennsylvania State University, State College 16802, Pennsylvania USA}

\begin{abstract}
 Quantum superintegrable systems are  solvable  eigenvalue problems. Their solvability is due to symmetry, but the symmetry is often ``hidden''.
The symmetry generators of  2nd order superintegrable systems in 2 dimensions close under commutation to define quadratic algebras, 
a generalization of Lie algebras. 
Distinct  systems
and their  algebras are related by geometric limits, induced by generalized In\"on\"u-Wigner  Lie algebra  contractions of the symmetry
algebras of the underlying spaces.
These  have  physical/geometric implications, such as the Askey scheme for  hypergeometric orthogonal polynomials.  
The systems can be best understood by  transforming them  to Laplace conformally superintegrable systems 
and using ideas  introduced in the 1894 thesis of B\^ocher to study separable solutions of the wave equation.  
The contractions can be subsumed into contractions of the conformal algebra $so(4,\C)$ to itself. 
Here we announce  main findings, with detailed classifications  in papers under preparation.
\end{abstract}

\keywords{conformal superintegrability, contractions, Laplace equations}

\maketitle

\section{Introduction}
A  quantum superintegrable system  is an integrable Hamiltonian system on an $n$-dimensional Riemannian/pseudo-Riemannian manifold 
with potential: $H=\Delta_n+V$   that admits $2n-1$ 
algebraically independent  partial differential operators $L_j$ commuting with $H$, the maximum  possible. 
$ [H,L_j]=0,\quad   \ j=1,2,\cdots, 2n-1$.
Superintegrability captures the properties of 
quantum Hamiltonian systems that allow the Schr\"odinger eigenvalue problem  (or Helmholtz equation)  $H\Psi=E\Psi$ to be solved exactly, analytically and algebraically, \cite{EVAN,TTW2001, SCQS, FORDY, MPW2013}.     
  A system is  of order $K$ if the maximum order of the symmetry
 operators, other than $H$, is  $K$. For $n=2$, $K=1,2$ all systems are known, e.g. \cite{KKM20041,DASK2007} 

 We review quickly the facts for {\it free} 2nd order superintegrable systems, (i.e., no potential, $K=2$) in the case $n=2, 2n-1=3$. The complex spaces with Laplace-Beltrami
 operators admitting at least three 2nd order symmetries were classified by Koenigs (1896), \cite{Koenigs}. They are:
\begin{itemize} \item The two constant curvature spaces (flat space and the complex sphere), six linearly independent 2nd order symmetries and three 1st order symmetries,
 \item The four  Darboux spaces (one with a parameter), four 2nd order symmetries and one 1st order symmetry,
\[ ds^2=4x(dx^2+dy^2),\ ds^2=\frac{x^2+1}{x^2}(dx^2+dy^2),\]
\[ds^2=\frac{e^x+1}{e^{2x}}(dx^2+dy^2),\ ds^2= \frac{2\cos 2x+b}{\sin^2 2x}(dx^2+dy^2),\] \cite{KKMW}
\item Eleven 4-parameter Koenigs spaces. No 1st order symmetries. An example is  
\[ ds^2=(\frac{c_1}{x^2+y^2}+\frac{c_2}{x^2}+\frac{c_3}{y^2}+c_4)(dx^2+dy^2).\]
\end{itemize}
For 
2nd order systems with non-constant potential, $K=2$, the following is true \cite{KKM20041,DASK2007,Zhedanov1992a,BDK,VILE}.
 \begin{itemize}
\item  The symmetry operators of 
each system close under commutation to generate a 
quadratic algebra, and 
the irreducible representations of this algebra determine the eigenvalues of $H$ and their multiplicity 
\item 
 All the 2nd order superintegrable systems are limiting cases of
 a single system: the generic 3-parameter potential on the 2-sphere,  $S_9$ in our listing, \cite{KKMP}, or are obtained from these limits by a St\"ackel 
transform (an invertible  structure preserving mapping of superintegrable systems, \cite{KKM20041}).  Analogously all  quadratic symmetry algebras of these 
systems are limits of that of $ S_9$.
\[ {S_9}:\qquad H=\Delta_2+\frac{a_1}{s_1^2}+\frac{a_2}{s_2^2}+\frac{a_3}{s_3^2}, \quad s_1^2+s_2^2+s_3^2=1,\]
\[L_1= (s_2\partial_{s_3}-s_3\partial_{s_2})^2 +\frac{a_3 s_2^2}{s_3^2}+\frac{a_2 s_3^2}{s_2^2},\ L_2,\ L_3,\]\item  2nd order superintegrable systems are multiseparable.
\end{itemize}
Here we consider only the { nondegenerate superintegrable systems}: Those with 4-parameter potentials (the maximum possible):
\[V({\bf x})= a_1V_{(1)}({\bf x})+a_2V_{(2)}({\bf x})+a_3V_{(3)}({\bf x})+a_4,\]
where $\{V_{(1)}({\bf x}),\ V_{(2)}({\bf x}),\ V_{(3)}({\bf x}),\ 1\}$ is a linearly independent set.
For these the symmetry algebra generated by $H,L_1,L_2$ always closes under commutation and gives the following quadratic algebra structure: 
Define  3rd order commutator $R$ by $ R=[L_1,L_2]$. Then  
\begin{small}
\[[L_j,R]=
A_1^{(j)}L_1^2+A_2^{(j)}L_2^2+A_3^{(j)}H^2+A_4^{(j)}\{L_1,L_2\}+A_5^{(j)}HL_1\]
\[+A_6^{(j)}HL_2 {+A_7^{(j)}L_1+A_8^{(j)}L_2+A_9^{(j)}H+A_{10}^{(j)}},\]
\[ R^2 = b_1 L_1^3 + b_2 L_2^3 + b_3 H^3 + b_4 \{L_1^2,L_2\} + b_5 \{L_1,L_2^2\}\]
\[ + b_6
L_1 L_2 L_1 + b_7 L_2 L_1 L_2
 + b_8 H\{L_1,L_2\} +b_9 H L_1^2 + b_{10} H L_2^2 \]
\[+ b_{11} H^2
L_1 + b_{12} H^2 L_2 + b_{13} L_1^2 + b_{14} L_2^2 + b_{15} \{L_1,L_2\} \]
\[+ b_{16} H L_1 + b_{17} H
L_2 + b_{18} H^2 + b_{19} L_1 + b_{20} L_2 + b_{21} H + b_{22},\] \end{small}
where $ \{L_1,L_2\}=L_1L_2+L_2L_1$ and the $A_i^{(j)},b_k$ are constants.

All 2nd order 2D  superintegrable systems with potential and their quadratic algebras are known. There are  { 44 } nondegenerate systems, on a variety of manifolds (just the manifolds 
classified by Koenigs), 
but under { the St\"ackel 
transform}  they divide into 6 equivalence classes with representatives on flat space and the 2-sphere, \cite{Kress2007}.
 Every 2nd order symmetry operator on a constant curvature space takes the form
\[ L= K + W({\bf x}),\]
where $K$ is a 2nd order element in the enveloping algebra of $o(3,\C)$ or $e(2,\C)$.
An example is  $S_9$ where
 \[ H=J_1^2+J_2^2+J_3^2+\frac{a_1}{s_1^2}+\frac{a_2}{s_2^2}+\frac{a_3}{s_3^2}\]
where      $J_3=s_1\partial_{s_2}-s_2\partial_{s_1}$  and $J_2,J_3$
are obtained by cyclic permutations of indices.  
{ Basis symmetries are} ($J_3=s_2\partial_{s_1}-s_1\partial_{s_2}, \cdots$)
\[L_1=J_1^2+\frac{a_3 s_2^2}{s_3^2}+\frac{a_2 s_3^2}{s_2^2},\
 L_2=J_2^2+\frac{a_1 s_3^2}{s_1^2}+\frac{a_3 s_1^2}{s_3^2},\]
\[ L_3=J_3^2+\frac{a_2 s_1^2}{s_2^2}+\frac{a_1 s_2^2}{s_1^2}.\]

 \begin{theorem}\label{theorem1} There is a bijection between quadratic algebras generated by 2nd order elements in the enveloping algebra of $o(3,\C)$, {called free}, and 2nd order nondegenerate superintegrable systems on the complex 2-sphere. Similarly, there is a bijection between quadratic algebras generated by 2nd order elements in the enveloping algebra of $e(2,\C)$ and 2nd order nondegenerate superintegrable systems on the 2D complex flat space.
  \end{theorem}
Remark :{ This theorem is constructive, \cite{KM2014}. Given a free quadratic algebra $\tilde Q$  one can compute the potential $V$ and 
the symmetries of the quadratic algebra $Q$ of the nondegenerate superintegrable system.}

 Special functions arise from these systems in  two distinct ways: 1)
 As  separable eigenfunctions of the quantum Hamiltonian.  Second order superintegrable systems are multiseparable, \cite{KKM20041}. 
 2) As  interbasis expansion coefficients relating distinct separable coordinate eigenbases, \cite{KMP2007a,KMP2008,P2011,LM2014}. 
 Most of the classical special functions in the Digital Library of Mathematical Functions, as well as Wilson polynomials, appear in these ways, \cite{DLMF}.

\subsection{The big picture: Contractions and special functions}

 \begin{itemize}\item Taking coordinate limits starting from  quantum system $S_9$ we can obtain other superintegrable systems.
  \item These coordinate limits induce limit relations between the special functions associated as eigenfunctions of  the superintegrable systems.
  \item The limits induce contractions of the associated quadratic algebras, and via the models, limit relations between the associated 
  special functions.
  \item For constant curvature systems the required limits are all induced by In\"on\"u-Wigner-type Lie algebra contractions of $o(3,\C)$ and $e(2,\C)$, \cite{Wigner,WW,NP}
  \item The Askey scheme for orthogonal functions of hypergeometric type fits nicely into this picture. \cite{KMP2014}
 \end{itemize}

 {Lie algebra contractions}:
Let $(A; [ ; ]_A)$, $(B; [ ; ]_B)$ be two complex Lie algebras. We say that
$B$ is a {\it contraction} of $A$ if for every $\epsilon\in (0; 1]$ there exists a linear invertible
map $t_\epsilon : B\to A$ such that for every $X, Y\in B$, 
$ \lim_{\epsilon\to 0}t_\epsilon^{-1}[t_\epsilon X,t_\epsilon Y]_A
= [X, Y ]_B$. 
{Thus,  as $\epsilon\to 0$ the 1-parameter family of basis transformations can become nonsingular but 
the structure constants go to a finite limit.}

{Contractions of $e(2,\C)$ and $o(3,\C)$:}
These are the  symmetry Lie algebras of free (zero potential) systems on constant curvature spaces.
Their contractions  have long since been classified, \cite{KM2014}. There are 6 nontrivial contractions of  $e(2,\C)$ and 4 of $o(3,\C)$. They are each induced by coordinate limits.

\medskip
{  Example: An In\"on\"u-Wigner- contraction of $o(3,\C)$.}
We use the classical realization for $o(3,\C)$ acting on the 2-sphere, with basis $J_1=s_2p_3-s_3p_2,\ J_2=s_3p_1-s_1p_3,\ J_3=s_1p_2-s_2p_1$, commutation relations
$ [J_2,J_1]=J_3,\quad [J_3,J_2]=J_1,\quad [J_1,J_3]=J_2$,
and Hamiltonian $H=J_1^2+J_2^2+J_3^2$. Here $s_1^2+s_2^2+s_3^2=1$. 
We introduce the basis change:
\[\{J_1',J_2',J_3'\}=\{ \epsilon J_1,\ \epsilon J_2,\  J_3\},\ 0<\epsilon\le 1,\] with coordinate 
 implementation $ x=\frac{s_1}{\epsilon},y=\frac{s_2}{\epsilon}, s_3\approx 1$.
The structure relations become \[  [ J_2',J_1']=\epsilon^2J_3',\  [J_3',J_2']=J_1',\ [J_1',J_3']=J_2',\]
As $\epsilon\to 0$ these converge to \[ [ J_2',J_1']=0,\  [J_3',J_2']=J_1',\ [J_1',J_3']=J_2',\] the Lie algebra $e(2,\C)$.

{Contractions of quadratic algebras:}
Just as for Lie algebras we can define a contraction of a quadratic algebra in terms of 1-parameter families of basis changes in the algebra: 
{ As $\epsilon\to 0$ the 1-parameter family of basis transformations  becomes singular but  the structure constants go to a finite limit,\cite{KM2014}.}

Motivating idea: {Lie algebra contractions induce quadratic algebra contractions. }
For constant curvature spaces we have
\begin{theorem}\cite{KM2014}, { Every Lie algebra contraction of $A=e(2,\C)$ or $A=o(3,\C)$  induces  a contraction of a free 
(zero potential) quadratic
 algebra $\tilde Q$ based on $A$,
which in turn induces  a contraction of the quadratic algebra $Q$ with potential. This is true for both classical and quantum algebras. }
\end{theorem}

\subsection{The problems and the proposed solutions}
The various limits of 2nd order superintegrable systems on constant curvature spaces and their applications, such as to the Askey-Wilson scheme, 
can be classified and understood in terms of generalized In\"on\"u-Wigner contractions \cite{KM2014}.
However, there are complications for spaces not of constant curvature.
For  Darboux spaces the Lie symmetry algebra is only 1-dimensional so
limits must be 
determined on a case-by-case basis. There is no Lie symmetry algebra at all for Koenigs spaces. Furthermore, there is the issue of finding a more systematic way of classifying the 
44 distinct Helmholtz superintegrable eigenvalue systems  on different manifolds, and their relations. 
These issues can be clarified by considering the Helmholtz systems as Laplace equations (with potential) on flat space. 
This point of view was introduced in the paper \cite{Laplace2011} and applied in \cite{CapelKress} to solve important classification problems in the case $n=3$.
It is the aim of this paper to describe the Laplace equation mechanism and how it can be applied to systematize the classification of Helmholtz 
superintegrable systems and their relations via limits. The basic idea is that families of (St\"ackel-equivalent) Helmholtz superintegrable systems
on a variety of manifolds correspond to a single conformally superintegrable Laplace equation on flat space. We exploit this 
relation in the case $n=2$,
but it generalizes easily to all dimensions $n\ge 2$. The conformal symmetry algebra for Laplace equations with constant potential on flat space is the
conformal algebra $so(n+2,\C)$. In his 1894 thesis \cite{Bocher} B\^ocher introduced a limit procedure based on the roots of quadratic forms to find families of 
R-separable solutions of the ordinary (zero potential) flat space Laplace equation in $n$ dimensions. We show that his limit procedure can be interpreted as constructing
generalized In\"on\"u-Wigner Lie algebra contractions of $so(4,\C)$ to itself. We call these B\^ocher contractions and show that all of the 
limits  of the
Helmholtz systems classified before for $n=2$, \cite{KM2014}, are induced by the larger class of B\^ocher contractions.
Here we present the main constructions and findings. Detailed proofs and the lengthy classifications  are in  papers under preparation.

\section{The Laplace equation}

Systems of  Laplace type are of the form 
$H\Psi\equiv \Delta_n\Psi+V\Psi=0$.
 Here $\Delta_n $ is the Laplace-Beltrami operator on a conformally flat nD Riemannian or pseudo-Riemannian manifold. 
 A conformal symmetry of this equation is a partial differential operator  $ S$ in the variables ${\bf x}=(x_1,\cdots,x_n)$  such 
that $[ S, H]\equiv SH-HS=R_{ S} H$ for some differential operator  $R_{S}$.
The system is maximally {\it conformally superintegrable} (or Laplace superintegrable) for $n\ge 2$  if there are $2n-1$ functionally independent conformal symmetries, 
${ S}_1,\cdots,{ S}_{2n-1}$ with ${ S}_1={ H}$, \cite{Laplace2011}. It is second order conformally superintegrable if each 
 symmetry $S_i$ can be chosen to be a  differential operator of at most second order.
Every $2D$ Riemannian manifold is conformally flat, so we can  always find a Cartesian-like coordinate system with coordinates $(x,y)\equiv (x_1,x_2)$ 
such that the Helmholtz eigenvalue  equation takes the form
\be\label{eqn1}
 {\tilde H}\Psi=\left(\frac{1}{\lambda(x,y)}(\partial_x^2+\partial_y^2)+{\tilde V}({\bf x})\right)\Psi=E\Psi.\ee
However, this equation is equivalent to the flat space Laplace  equation
\be\label{eqn2} { H}\Psi\equiv \left(\partial_x^2+\partial_y^2+ V({\bf x})\right)\Psi=0,\ V({\bf x})=\lambda({\bf x})({\tilde V}({\bf x})-E).\ee
In particular, the  symmetries of (\ref{eqn1})  correspond to  the conformal symmetries of (\ref{eqn2}). Indeed, if $[S,{\tilde H}]=0$ then 
\[ [S,H]=[S,\lambda({\tilde H}-E)]=[S,\lambda]({\tilde H}-E)=[S,\lambda]\lambda^{-1}H.\]
Conversely,  if $S$ is an $E$-independent conformal symmetry of $H$ we find that $[S,{\tilde H}]=0$. Further, the conformal symmetries of the system
$({\tilde H}-E)\Psi=0$ are identical with the conformal symmetries of (\ref{eqn2}).
Thus without loss of generality  we can assume the manifold is flat space with $\lambda\equiv 1$. 

\medskip

{\bf The conformal St\"ackel transform}:
Suppose we have a second order conformal  superintegrable system 
\[ { H}=\partial_{xx}+\partial_{yy}+V(x,y)=0,\quad  { H}={ H}_0+V
\]
where $V(x,y)= W(x,y)-E\ U(x,y)$ for arbitrary parameter $E$. 
\begin{theorem}: The transformed (Helmholtz) system  ${\tilde {  H}}\Psi=E\Psi,\quad 
{\tilde { H}}=\frac{1}{{ U}}(\partial_{xx}+\partial_{yy})+{\tilde V}
$
is truly  superintegrable, where
  ${\tilde V}=\frac{W}{U}$, \cite{Laplace2011}.\end{theorem}
There is a similar definition of ordinary St\"ackel transforms of  Helmholtz superintegrable systems $H\Psi=E\Psi$  which takes superintegrable systems to superintegrable systems, essentially preserving the quadratic algebra structure, \cite{CCM}.

Thus  any second order conformal Laplace superintegrable system admitting a nonconstant potential $U$ can be 
St\"ackel transformed to a Helmholtz
superintegrable system. 
By choosing all possible special potentials $U$ associated with the  fixed Laplace system  we generate the equivalence class of all Helmholtz 
superintegrable systems  obtainable through this process. 
 \begin{theorem} {There is a one-to-one relationship between 
 flat space  conformally superintegrable Laplace systems with nondegenerate potential   and St\"ackel equivalence classes of superintegrable Helmholtz systems with nondegenerate potential. } \end{theorem}

Indeed,  for a St\"ackel transform induced by the function 
 $U^{(1)}$, we can take the original Helmholtz system to have Hamiltonian
 \be\label{parameter} H=H_0+V=H_0+U^{(1)}\alpha_1+U^{(2)}\alpha_2+U^{(3)}\alpha_3+\alpha_4\ee
 where $\{U^{(1)},U^{(2)},U^{(3)},1\}$ is a basis for the 4-dimensional potential space. A 2nd order symmetry $S$ would have the form
 \[ S=S_0+W^{(1)}\alpha_1+W^{(2)}\alpha_2+W^{(3)}\alpha_3,\]
where $S_0$ is a symmetry of the potential free Hamiltonian, $H_0$.
 The St\"ackel transformed symmetry and Hamiltonian  take the form ${\tilde S}=S-W^{(1)}{\tilde H}$ and
 \[ {\tilde H}=\frac{1}{U^{(1)}}H_0+\frac{U^{(1)}\alpha_1+U^{(2)}\alpha_2+U^{(3)}\alpha_3+\alpha_4}{U^{(1)}}
.\]
 Note that the parameter $\alpha_1$ cancels out of the expression for $\tilde S$; it is replaced by a term $-\alpha_4W^{(1)}/U^{(1)}$. Now suppose that 
 $\Psi$ is a formal  eigenfunction of 
 $H$ (not required to be normalizable): $H\Psi=E\Psi$. Without loss of generality we can absorb the energy eigenvalue into $\alpha_4$ so that  $\alpha_4=-E$ in (\ref{parameter}) and,  in 
 terms of this redefined $H$, we have $H\Psi =0$.
 It follows immediately
 that ${\tilde S}\Psi =S\Psi$.  Thus, for the 3-parameter system $H'$ and the St\"ackel transform ${\tilde H}'$,
  \[H'=H_0+V'=H_0+U^{(1)}\alpha_1+U^{(2)}\alpha_2+U^{(3)}\alpha_3,\]
  \[{\tilde H}'=\frac{1}{U^{(1)}}H_0
  +\frac{-U^{(1)}E+U^{(2)}\alpha_2+U^{(3)}\alpha_3}{U^{(1)}},\]
we have $H'\Psi=E\Psi $ and ${\tilde H}'\Psi=-\alpha_1\Psi$. The effect of the St\"ackel transform is 
to replace $\alpha_1$ by $-E$ and $E$
by $-\alpha_1$. Further,  $S$ and $\tilde S$  must agree on eigenspaces of $H'$.

We know that the symmetry operators of all 2nd order nondegenerate superintegrable systems in 2D generate a quadratic algebra
of the form 
\[{} [R,S_j]=f^{(j)}(S_1,S_2,\alpha_1,\alpha_2,\alpha_3,H'),\ j=1,2,\]
\be\label{quadratic1} R^2=f^{(3)}(S_1,S_2,\alpha_1,\alpha_2,\alpha_3,H'),\ee
where $\{S_1,S_2,H\}$ is a basis for the 2nd order symmetries and $\alpha_1,\alpha_2,\alpha_3$ are the parameters 
for the potential, \cite{KKM20041}.  It follows from the above considerations that the effect of a St\"ackel transform generated by the 
potential function $U^{(1)}$ is to determine a new superintegrable 
system with structure 
\[{} [{\tilde R},{\tilde S}_j]=f^{(j)}({\tilde S}_1,{\tilde S}_2,-{\tilde H}',\alpha_2,\alpha_3,-\alpha_1),\ j=1,2,\]
\be\label{quadratic2} R^2=f^{(3)}({\tilde S}_1,{\tilde S}_2,-{\tilde H}',\alpha_2,\alpha_3,-\alpha_1),\ee
Of course, the switch of $\alpha_1$ and $H'$ is only for illustration; there is a St\"ackel transform that replaces any 
$\alpha_j$ by $-H'$ and $H'$ by $-\alpha_j$ and similar transforms that apply to any basis that we choose for the potential space. 

Formulas (\ref{quadratic1}) and (\ref{quadratic2}) are just instances of the quadratic algebras of the superintegrable systems belonging to the 
equivalence class of a single nondegenerate conformally superintegrable Hamiltonian
\be\label{confham}\hat{H}=\partial_{xx}+\partial_{yy}+\sum_{j=1}^4\alpha_j V^{(j)}(x,y).\ee
Let $\hat{S}_1,\hat{S}_2, \hat{H}$ be a basis of 2nd order conformal  symmetries of $\hat H$. From the above discussion we can conclude the following.
\begin{theorem} The  symmetries of the 2D nondegenerate conformal superintegrable Hamiltonian $\hat H$ generate a quadratic algebra
 \be\label{confquadalg} [{\hat R},{\hat S}_1]=f^{(1)}({\hat S}_1,\hat{S}_2,\alpha_1,\alpha_2,\alpha_3,\alpha_4),\ee
 \[[{\hat R},{\hat S}_2]=f^{(2)}
 ({\hat S}_1,{\hat S}_2,\alpha_1,\alpha_2,\alpha_3,\alpha_4),\]
\[ {\hat R}^2=f^{(3)}({\hat S}_1,\hat{S}_2,\alpha_1,\alpha_2,\alpha_3,\alpha_4),\]
where $\hat{R}=[{\hat S}_1,\hat{S}_2]$ and all identities hold ${\rm mod}({\hat H})$. A conformal St\"ackel transform generated by the potential 
$V^{(j)}(x,y)$ yields a nondegenerate Helmholtz superintegrable Hamiltonian $\tilde H$ with quadratic algebra relations identical to (\ref{confquadalg}),
except that we make the replacements ${\hat S}_\ell\to {\tilde  S}_\ell$ for $\ell=1,2$   and $\alpha_j\to -{\tilde H}$. These modified relations  
(\ref{confham}) are now true identities, not ${\rm mod}({\hat H})$.
\end{theorem}

Every 2nd order conformal symmetry is of the form $S=S_0+W$ where $S_0$ is a 2nd order element of the enveloping algebra of $so(4,\C)$.
The dimension of this space of 2nd order elements is 21 but there is an 11-dimensional subspace of symmetries congruent to 0 ${\rm mod} (H_0)$ 
where $H_0=P_1^2+P_2^2$. 
Thus ${\rm mod} (H_0)$ the space of 2nd order symmetries is 10-dimensional.

\section{The B\^ocher Method}
 In his 1894 thesis B\^ocher, \cite{Bocher}, developed a geometrical method for 
finding and classifying the R-separable orthogonal coordinate systems for the flat space Laplace equation $\Delta_n\Psi=0$ in $n$ dimensions. It was based on the conformal
symmetry of these equations. The conformal Lie symmetry algebra of the flat space   complex Laplacian  is $so(n+2,\C)$. We will use his ideas for $n=2$ , but applied to the Laplace equation with potential 
$H\Psi\equiv (\partial^2_x+\partial^2_y+V)\Psi=0$.
The $so(4,\C)$  conformal symmetry algebra in the case $n=2$ has the basis
$P_1=\partial_x,\ P_2=\partial_y,\  J=x\partial_y-y\partial_x,\
D=x\partial_x+y\partial_y,\
K_1=(x^2-y^2)\partial_x +2xy\partial_y,\ K_2=(y^2-x^2)\partial_y+2xy\partial_x
$.
B\^ocher linearizes this action by introducing tetraspherical coordinates. These are 4 projective complex coordinates $(x_1,x_2,x_3,x_4)$ 
confined to the null cone $x_1^2+x_2^2+x_3^2+x_4^2=0$.
They are related to complex  Cartesian coordinates $(x,y)$ via
\[ x=-\frac{x_1}{x_3+ix_4},\quad  y=-\frac{x_2}{x_3+ix_4},\]  
\[  H=\partial_{xx}+\partial_{yy}+{\tilde V}=(x_3+ix_4)^2\left(\sum_{k=1}^4\partial_{x_k}^2+V\right)\]
where ${\tilde V}=(x_3+ix_4)^2V$.
We define
$ L_{jk}=x_j\partial_{x_k}-x_k \partial_{x_j}, \ 1\le j,k\le 4,\ j\ne k$,
where $L_{jk}=-L_{kj}$. These operators are clearly a  basis for $so(4,\C)$. The generators for flat space conformal 
symmetries are related to these via
\[P_1= \partial_x=L_{13}+iL_{14},\ P_2=\partial_y=L_{23}+iL_{24},\  D=iL_{34},\]
\[ J=L_{12},\  K_1=L_{13}-iL_{14},\  K_2=L_{23}-iL_{24}.\]

\subsection{Relation to separation of variables}
B\^ocher uses symbols of the form $[n_1,n_2,..,n_p]$ where 
$n_1+...+n_p=4$,  to define coordinate surfaces as 
follows. Consider the quadratic forms
\[ \Omega =x^2_1+x^2_2+x^2_3+x^2_4=0,\]  
\[\Phi =\frac{x^2_1}{ \lambda -e_1} + \frac{x^2_2}{ \lambda -e_2} + 
\frac{x^2_3}{ \lambda -e_3} + \frac{x^2_4}{ \lambda -e_4}.\]
If the parameters $e_j$ are pairwise distinct, the elementary divisors of these two  forms are denoted by  
$[1,1,1,1]$.
 Given a point in 2D flat space with Cartesian coordinates $(x^0,y^0)$, there corresponds a set of tetraspherical coordinates 
$(x^0_1,x^0_2,x^0_3,x^0_4)$, unique up to multiplication by a nonzero constant. If we  substitute into $\Phi$
we see that there are exactly 2 roots $\lambda=\rho,\mu$ such that $\Phi=0$. 
(If $e_4\to\infty$ these correspond to elliptic coordinates on the 2-sphere.) They are orthogonal with respect to the metric $ds^2=dx^2+dy^2$ and  are $R$-separable for the 
Laplace equations $(\partial^2_x+\partial^2_y)\Theta=0$ or $(\sum_{j-1}^4\partial_{x_j}^2)\Theta=0$. 
\begin{example}
Consider the  potential 
$V_{[1,1,1,1]}=\frac{a_1}{ x^2_1} + \frac{a_2}{ x^2_2} + 
\frac{a_3}{ x^2_3} + \frac{a_4}{ x^2_4}$.
It is the only  potential $V$ such that  equation $(\sum_{j-1}^4\partial_{x_j}^2+V)\Theta=0$ is $R$-separable in elliptic 
coordinates for {\it all} choices of  the parameters $e_j$. The separation is characterized by 2nd order conformal symmetry operators that are 
linear in the parameters 
$e_j$. In particular the symmetries span a  3-dimensional subspace of symmetries, so the system $H\Theta=(\sum_{j=1}^4\partial_{x_j}^2+V_{[1,1,1,1]})\Theta=0$ must be conformally 
superintegrable.\end{example} 

\subsection{B\^ocher limits}

 Suppose some of the $e_i$ become equal. To obtain separable coordinates we cannot just set them equal in $\Omega,\Phi$  but must take limits, 
  B\^ocher develops a calculus to describe this. Thus the process of 
making $e_1\rightarrow e_2$ is described by the mapping, which in the limit as $\epsilon\to 0$ takes the null cone to the null cone. 
\[ e_1=e_2+\epsilon ^2,\  
x_1\rightarrow \frac{i(x'_1+ix'_2)}{ \sqrt{2}\epsilon },\]
\[x_2\rightarrow \frac{(x'_1+ix'_2)}{\sqrt{2} \epsilon } + \epsilon\frac{ (x'_1-ix'_2)}{\sqrt{2}}, \ 
x_j\rightarrow x'_j,j=3,4,\]
In the limit we have \[\Omega={x_1'}^2+{x_2'}^2+{x_3'}^2+{x_4'}^2=0,\]
\[ \Phi=\frac{(x_1'+ix_2')^2}{2(\lambda-e_2)^2}+\frac{{x_1'}^2+{x_2'}^2}{\lambda-e_2}+\frac{{x_3'}^2}{ \lambda -e_3} +
\frac{{x_4'}^2}{ \lambda -e_4},\]
which has elementary divisors $[2,1,1]$, \cite{Proceedings, Bromwich}. In the same way as for $[1,1,1,1]$, these forms define a new set of orthogonal 
coordinates  $R$-separable for the Laplace equations. 
We can show that the coordinate limit induces a contraction of  $so(4,\C)$ to itself:
 \[ L'_{12}=L_{12},\ L'_{13}=-\frac{i}{\sqrt{2}\ \epsilon}(L_{13}-iL_{23})-\frac{i\ \epsilon}{\sqrt{2}}L_{13},\]
\[ L'_{23}=-\frac{i}{\sqrt{2}\ \epsilon}(L_{13}-iL_{23})-\frac{\ \epsilon}{\sqrt{2}}L_{13},\ 
    L'_{34}=L_{34},\]
    \[L'_{14}=-\frac{i}{\sqrt{2}\ \epsilon}(L_{14}-iL_{24})-\frac{i\ \epsilon}{\sqrt{2}}L_{14},\]
\[ L'_{24}=-\frac{i}{\sqrt{2}\ \epsilon}(L_{14}-iL_{24})-\frac{\ \epsilon}{\sqrt{2}}L_{14}.\]
 We call this  the B\^ocher contraction $[1,1,1,1]\to [2,1,1]$. There are analogous B\^ocher contractions of $so(4,\C)$ to 
itself corresponding to limits from $[1,1,1,1]$ to 
$[2,2],[3,1],[4]$. Similarly, there are B\^ocher contractions $[2,1,1]\to [2,2]$, etc. 

If we apply the contraction $[1,1,1,1]\to [2,1,1]$ to the potential $V_{[1,1,1,1]}$ we get a finite limit
\be \label{V[211]} V_{[2,1,1]}=\frac{b_1}{(x'_1+ix'_2)^2}+\frac{b_2(x'_1-ix'_2)}{(x'_1+ix'_2)^3}+\frac{b_3}{{x'_3}^2}
+\frac{b_4}{{x'_4}^2},\ee
provided the parameters transform as  
\[ a_1=-\frac12(\frac{b_1}{\epsilon^2}+\frac{b_2}{2\epsilon^4}),\ a_2=- \frac{b_2}{4\epsilon^4},\ a_3=b_3,\ a_4=b_4.\]
Note: We know from theory that the 4-dimensional vector space of potentials $V_{[1,1,1,1]}$ maps to the 4-dimensional vector space of potentials $V_{[2,1,1]}$ 1-1 under the contraction, \cite{KM2014}.
The reason for the $\epsilon$-dependence of the parameters is the arbitrariness of choosing a basis. If we had chosen a basis for $V_{[1,1,1,1]}$ specially adapted to this contraction, we could have achieved $a_j=b_j,\ 1\le j\le 4$.

B\^ocher contractions obey a composition law:
\begin{theorem}Let 
\[ A:  \left(\Delta_{\bf x}+V_A({\bf x})\right)\Psi=0,\quad B: 
 \left(\Delta_{\bf y}+V_B({\bf y})\right)\Psi=0,\]
\[\quad C: 
\left(\Delta_{\bf z}+V_C({\bf z})\right)\Psi=0,\]
be B\^ocher superintegrable systems such that $A$ B\^ocher-contracts to $B$ and $B$ B\^ocher-contracts to $C$. Then there is a one-parameter  contraction of $A$ to $C$.\end{theorem}

A fundamental advantage in recognizing B\^ocher's limit procedure as contractions is that whereas the B\^ocher limits had a fixed starting and ending point, say $[1,1,1,1]\to [2,1,1]$, contractions can be applied to any nondegenerate conformally superintegrable system and are guaranteed to result in another 
nondegenerate conformally superintegrable system. This greatly increases the range of applicability of the limits.

\section{The 8 classes of  nondegenerate conformally superintegrable systems}
The possible Laplace equations (in tetraspherical coordinates) are 
$(\sum_{j=1}^4\partial_{x_j}^2+V)\Psi=0$ with potentials:
\bea V_{[1,1,1,1]}&=&\sum_{j=1}^4\frac{a_j}{x_j^2},\label{eqiuivclasses}\\
V_{[2,1,1]}&=&\frac{a_1}{x_1^2}+\frac{a_2}{x_2^2}+\frac{a_3(x_3-ix_4)}{(x_3+ix_4)^3}+\frac{a_4}{(x_3+ix_4)^2},\nonumber\\
 V_{[2,2]}&=&\frac{a_1}{(x_1+ix_2)^2}+\frac{a_2(x_1-ix_2)}{(x_1+ix_2)^3}
+\frac{a_3}{(x_3+ix_4)^2}\nonumber\\
&+&\frac{a_4(x_3-ix_4)}{(x_3+ix_4)^3},\nonumber\\
V_{[3,1]}&=&\frac{a_1}{(x_3+ix_4)^2}+\frac{a_2x_1}{(x_3+ix_4)^3}\nonumber\\
&+&\frac{a_3(4{x_1}^2+{x_2}^2)}{(x_3+ix_4)^4}+\frac{a_4}{{x_2}^2},\nonumber\\
V_{[4]}&=&\frac{a_1}{(x_3+ix_4)^2}+a_2\frac{x_1+ix_2}{(x_3+ix_4)^3}\nonumber\\
&+&a_3\frac{3(x_1+ix_2)^2-2(x_3+ix_4)(x_1-ix_2)}{(x_3+ix_4)^4},\nonumber\\
 V_{[0]}&=&\frac{a_1}{(x_3+ix_4)^2}+\frac{a_2x_1+a_3x_2}{(x_3+ix_4)^3}\nonumber\\
&+&a_4\frac{x_1^2+x_2^2}{(x_3+ix_4)^4}.\nonumber\\
V(1)&=&a_1\frac{1}{(x_1+ix_2)^2}+a_2\frac{1}{(x_3+ix_4)^2}\nonumber\\
&+&a_3\frac{(x_3+ix_4)}{(x_1+ix_2)^3}
+a_4\frac{(x_3+ix_4)^2}{(x_1+ix_2)^4},\nonumber\\
V(2)&=&a_1\frac{1}{(x_3+ix_4)^2}+a_2\frac{(x_1+ix_2)}{(x_3+ix_4)^3}\nonumber\\
&+&a_3\frac{(x_1+ix_2)^2}{(x_3+ix_4)^4}
+a_4\frac{(x_1+ix_2)^3}{(x_3+ix_4)^5}.\nonumber\eea
(The last 3 systems do not correspond to elementary divisors; they appear as B\^ocher contractions of systems that do correspond to elementary divisors.) Each of the 44 Helmholtz nondegenerate superintegrable (i.e. 3-parameter) eigenvalue systems is St\"ackel equivalent to exactly one of these systems.  Thus, with one caveat,  there are exactly 8 equivalence classes of Helmholtz systems. The caveat is the singular family of  systems with potentials $V_S=(x_3+ix_4)^{-2}h(\frac{x_1+ix_2}{x_3+ix_4})$ where $h$ is an arbitrary analytic function
except that  $V_S\ne V(1),V(2)$. This family is unrelated to the other systems.

Expressed as flat space Laplace equations $(\partial_{x}^2+\partial_y^2+{\tilde V})\Psi=0$ in Cartesian coordinates, the potentials are
\bea 
 {\tilde V}_{[1,1,1,1]}&=&\frac{a_1}{x^2}+\frac{a_2}{y^2}+\frac{4a_3}{(x^2+y^2-1)^2}\nonumber\\
&-&\frac{4a_4}{(x^2+y^2+1)^2},\nonumber\\
{\tilde  V}_{[2,1,1]}&=&\frac{a_1}{x^2}+\frac{a_2}{y^2}-a_3(x^2+y^2)+a_4,\nonumber\\
{\tilde  V}_{[2,2]}&=&\frac{a_1}{(x+iy)^2}+\frac{a_2(x-iy)}{(x+iy)^3}\nonumber\\
&+&a_3-a_4(x^2+y^2),\nonumber\\
 {\tilde V}_{[3,1]}&=&a_1-a_2x +a_3(4x^2+{y}^2)+\frac{a_4}{{y}^2},\nonumber\\
 {\tilde V}_{[4]}&=&a_1-a_2(x+iy)\nonumber\\ 
&+&a_3\left(3(x+iy)^2+2(x-iy)\right)\nonumber\\
&-&a_4\left(4(x^2+y^2)+2(x+iy)^3\right),\nonumber\\
 {\tilde V}_{[0]}&=&a_1-(a_2x+a_3y)+a_4(x^2+y^2),\nonumber\\
{\tilde V}(1)&=&\frac{a_1}{(x+iy)^2}+a_2
-\frac{a_3}{(x+iy)^3}+\frac{a_4}{(x+iy)^4},\nonumber\\
 {\tilde V}(2)&=&a_1+a_2(x+iy)
+a_3(x+iy)^2\nonumber\\
&+&a_4(x+iy)^3.\label{Vcartesian}\eea

\subsection{Summary of B\^ocher contractions of Laplace superintegrable systems} \begin{enumerate}
\item{$[1,1,1,1]\to [2,1,1]$ contraction}: \[V_{[ [1,1,1,1]}\rightarrow V_{ [2,1,1]},\quad V_{[2,1,1]}\rightarrow V_{[2,1,1]},\]
\[ V_{[2,2]}\rightarrow V_{[2,2]},\quad
V_{[3,1]}\rightarrow V_{[2,1,1]},\]
\[ V_{[4]}\rightarrow V_{[0]},\  V_{[0]}\rightarrow V_{[0]},\  V(1)\rightarrow V(1),\   V(2)\rightarrow V(2). \]

\item{$[1,1,1,1]\to [2,2]$ contraction}: \[ V_{[1,1,1,1]}\rightarrow V_{[2,2]},\quad V_{[2,1,1]}\rightarrow V_{[2,2]},\]
\[ V_{[2,2]}\rightarrow V_{[2,2]},\quad
V_{[3,1]}\rightarrow V(1),\]
\[ V_{[4]}\rightarrow V(2),\  V_{[0]}\rightarrow V_{[0]},\   V(1)\rightarrow V(1),\   V(2)\rightarrow V(2). \]

\item{$[2,1,1]\to [3,1]$ contraction}: \[ V_{[1,1,1,1]}\rightarrow V_{[3,1]},\quad V_{[2,1,1]}\rightarrow V_{[3,1]},\]
\[ V_{[2,2]}\rightarrow V_{[0]},\quad
V_{[3,1]}\rightarrow V_{[3,1]},\]
\[ V_{[4]}\rightarrow V_{[0]},\ V_{ [0]}\rightarrow V_{[0]},\   V(1)\rightarrow V(2),\   V(2)\rightarrow V(2). \]

\item{$[1,1,1,1]\to [4]$ contraction}: \[ V_{[1,1,1,1]}\rightarrow V_{[4]},\quad V_{[2,1,1]}\rightarrow V_{[4]},\]
\[ V_{[2,2]}\rightarrow V_{[0]},\quad
V_{[3,1]}\rightarrow V_{[4]},\]
\[ V_{[4]}\rightarrow V_{[0]},\ V_{ [0]}\rightarrow V_{[0]},\   V(1)\rightarrow V(2),\   V(2)\rightarrow V(2). \]

\item{$[2,2]\to [4]$ contraction}: \[ V_{[1,1,1,1]}\rightarrow V_{[4]},\quad V_{[2,1,1]}\rightarrow V_{[4]},\]
\[V_{ [2,2]}\rightarrow V_{[4]},\quad
V_{[3,1]}\rightarrow V(2),\]
\[V_{ [4]}\rightarrow V(2),\  V_{[0]}\rightarrow V_{[0]},\   V(1)\rightarrow V(2),\   V(2)\rightarrow V(2). \]

\item{$[1,1,1,1]\to [3,1]$ contraction}: \[V_{ [1,1,1,1]}\rightarrow V_{ [3,1]},\quad V_{[2,1,1]}\rightarrow V_{[3,1]},\]
\[V_{ [2,2]}\rightarrow V_{ [3,1]},\quad
V_{[3,1}]\rightarrow V_{ [3,1]},\]
\[ V_{[4]}\rightarrow V_{[0]},\  V_{[0]}\rightarrow V_{ [0]},\   V(1)\rightarrow V(2),\quad   V(2)\rightarrow V(2). \]
\end{enumerate}
We have omitted some contractions, such as $[3,1]\to [4]$, because they are consequences of other contractions in the table.

\section{ Helmholtz contractions from B\^ocher contractions} We describe how B\^ocher contractions of conformal superintegrable systems induce contractions 
of Helmholtz  superintegrable systems.

We consider the conformal St\"ackel transforms of the conformal system $[1,1,1,1]$ with potential $V_{[1,1,1,1]}$. 
As we will show explicitly in another paper, the various possibilities are $S_9$ above and 2 more Helmholtz systems on the sphere, $S_7$ and 
$S_8$, 2 Darboux systems D4B and D4C, and a family of Koenigs systems.  
\begin{example}Using Cartesian coordinates
$x,y$, we consider the $[1,1,1,1]$ Hamiltonian
\[ H=\partial^2_x+\partial^2_y+ \frac{a_1}{ x^2} + \frac{a_2}{ y^2} + \frac{4a_3}{ (x^2+y^2-1)^2} + 
\frac{4a_4}{ (x^2+y^2+1)^2}.\]
Dividing  on the left by 
$1/x^2$ we obtain 
\[\hat H=x^2(\partial^2_x+\partial^2_y)+a_1 + a_2 \frac{x^2}{ y^2}+ 4a_3\frac{x^2}{ (x^2+y^2-1)^2}\]
\[- 4a_4\frac{x^2}{ (x^2+y^2+1)^2},\]
 the St\"ackel transform corresponding to the case  $(a_1,a_2,a_3,a_4)=(1,0,0,0)$. This becomes more transparent if we introduce variables 
$x=e^{-a},y=r$. The Hamiltonian $\hat H$  can be written 
\[\hat H=\partial^2_a+e^{-2a}\partial^2_r + a_1+ a_2\frac {e^{-2a}}{ r^2}
+
a_3 \frac{4}{ (e^{-a}+e^a(r^2-1))^2}\]
\[ - a_4\frac{4}{ (e^{-a}+e^a(r^2+1))^2}.\]
Recalling horospherical coordinates on the complex two sphere,
viz. 
\[s_1=\frac{i}{ 2}(e^{-a}+(r^2+1)e^a),\ 
s_2=re^a,\]
\[s_3=\frac{1}{ 2}(e^{-a}+(r^2-1)e^a)\]
we see that the Hamiltonian $\hat H$ can be written as  
\[ \hat H=\partial^2_{s_1}+ \partial^2_{s_2}+ \partial^2_{s_3}+ a_1+ \frac{a_2}{ s^2_2} + 
\frac{a_3}{ s^2_3} + \frac{a_4}{ s^2_1},\]
and this is explicitly the superintegrable system $S_9$.\end{example}

More generally, 
let $H$ be the initial Hamiltonian. In
terms of tetraspherical coordinates  a general conformal St\"ackel transformed potential will 
take the form 
\[ V=\frac{\frac{a_1}{x_1^2}+\frac{a_2}{x_2^2}+\frac{a_3}{x_3^2}+\frac{a_4}{x_4^2}}{\frac{A_1}{x_1^2}+\frac{A_2}{x_2^2}+\frac{A_3}{x_3^2}+\frac{A_4}{x_4^2}}
=\frac{V_{[1,1,1,1]}}{F({\bf x},{\bf A})},\]
where
\[ F({\bf x},{\bf A})=\frac{A_1}{x_1^2}+\frac{A_2}{x_2^2}+\frac{A_3}{x_3^2}+\frac{A_4}{x_4^2},\]
and the transformed Hamiltonian will be 
\[{\hat H}=\frac{1}{ F({\bf x},{\bf A})}H,\]
where the transform is determined by the fixed vector $(A_1,A_2,A_3,A_4)$. Now we apply the B\^ocher contraction $[1,1,1,1]\to [2,1,1]$ to this system.
In the limit as $\epsilon\to 0$ 
the potential $V_{[1,1,1,1]}\to V_{[2,1,1]}$, (\ref{V[211]}), and $H\to H'$ the $[2,1,1]$ system. Now consider
\[ F({\bf x}(\epsilon),{\bf A})= V'({\bf x}',A)\epsilon^\alpha+O(\epsilon^{\alpha+1}),\]
where the  the integer exponent $\alpha$ depends upon our choice of $\bf A$. We will provide the theory to show that the system defined by Hamiltonian
\[ {\hat H}'=\lim_{\epsilon\to 0}\epsilon^\alpha {\hat H}(\epsilon)=\frac{1}{V'({\bf x}',A)}H'\]
is a superintegrable system that arises from the  system $[2,1,1]$ by a conformal St\"ackel transform induced by the potential $V'({\bf x}',A)$.
Thus the Helmholtz superintegrable system with potential $V=V_{[1,1,1,1]}/F$ contracts to the Helmholtz superintegrable system with potential $V_{[2,1,1]}/V'$.
The contraction is induced by a generalized In\"on\"u-Wigner Lie algebra contraction of the conformal algebra $so(4,\C)$. Always  the $V'$ can be identified with a specialization of the
$[2,1,1]$ potential . Thus a conformal St\"ackel transform of $[1,1,1,1]$ has been contracted to a conformal St\"ackel transform of $[2,1,1]$. The results  follow and generalize to all Laplace systems.  
The basic idea  is that the procedure of taking a conformal St\"ackel transform of a conformal superintegrable system, followed
by a Helmholtz contraction yields the same result as taking a B\^ocher contraction followed by an ordinary St\"ackel transform: The 
diagrams commute. The possible Helmholtz contractions obtainable from these B\^ocher contractions number well over 100; they will be classified in another paper.

\begin{figure}[H]
\centering
\includegraphics[scale=0.5]{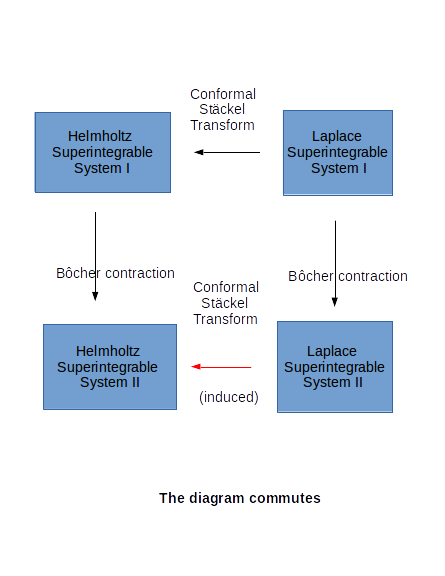}
\caption{Relationship between conformal St\"ackel transforms and B\^ocher contractions}
\end{figure}

{All quadratic algebra contractions are induced by Lie algebra contractions of $so(4,\C)$, even those for Darboux and Koenigs spaces. }

\begin{figure}[H]
\centering
\includegraphics[scale=0.08]{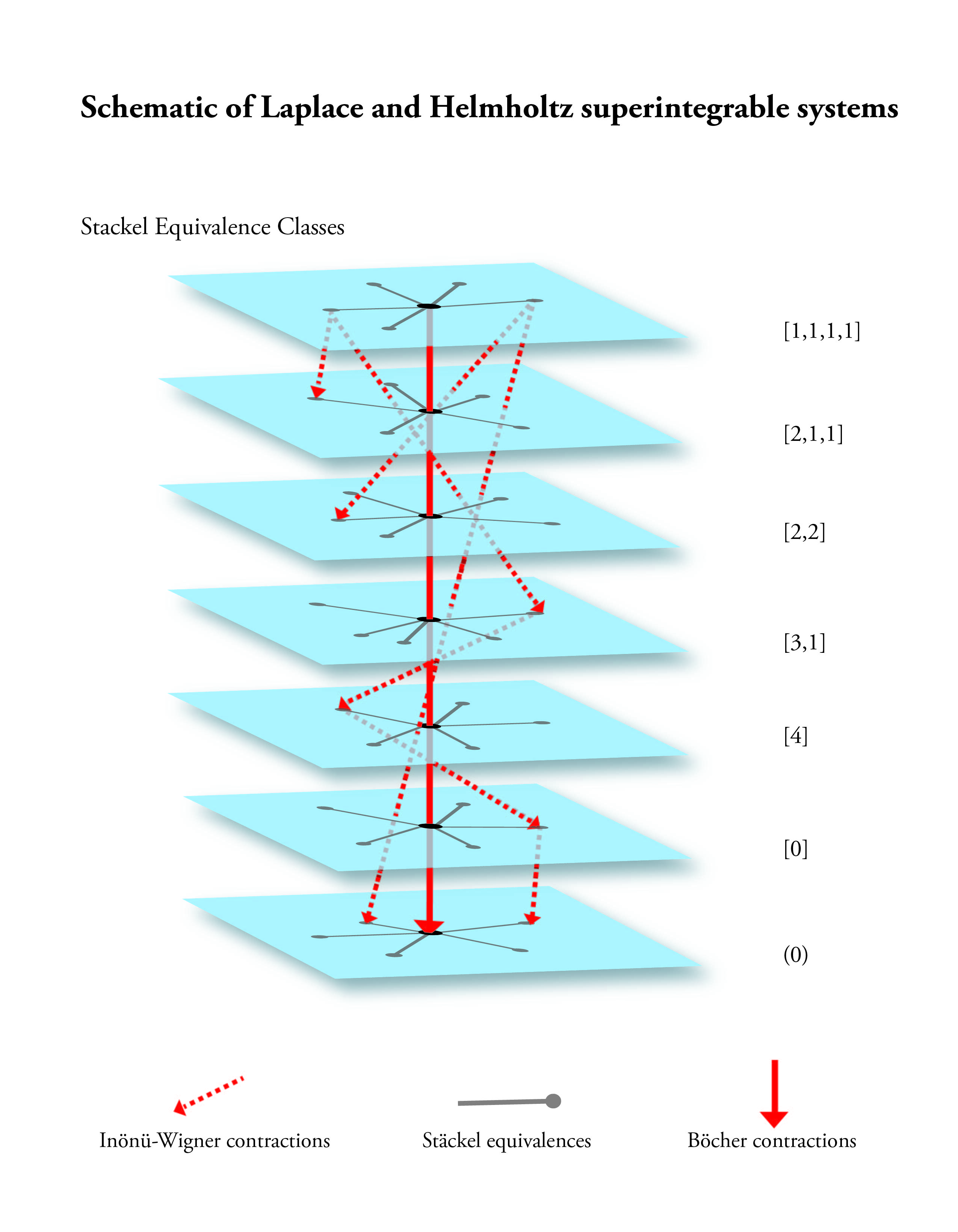}
\caption{The bigger picture}
\end{figure}

\section{Conclusions and discussion}

We have pointed out that the use of Lie algebra contractions based on the symmetry groups of constant curvature spaces to construct quadratic algebra contractions of 2nd order 2D Helmholtz superintegrable systems is incomplete, because it doesn't satisfactorily account for Darboux and Koenigs spaces, and because even for constant curvature spaces there are abstract quadratic algebra contractions that cannot be obtained from the Lie symmetry algebras. However, this gap is filled in when one extends these systems to 2nd order Laplace conformally superintegrable systems with conformal symmetry algebra. Classes of St\"ackel equivalent Helmholtz superintegrable systems are now recognized as corresponding to a single Laplace superintegrable system on flat space with underlying conformal symmetry algebra $so(4,\C)$.  
The conformal Lie algebra contractions are induced by  B\^ocher limits associated with invariants of quadratic forms. 
They generalize all of the Helmholtz contractions derived earlier.  In particular, contractions of Darboux and Koenigs systems can be 
described easily. All of the concepts introduced in this paper are clearly also applicable for dimensions $n\ge 3$, \cite{CKP2015}.

In a paper under preparation we will  1) give a complete detailed classification of 2D nondegenerate 2nd order conformally superintegrable systems and their relation to B\^ocher contractions, 2) present a detailed classification of   all B\^ocher contractions of  2D nondegenerate 2nd order conformally superintegrable systems., 3) present tables describing the  contractions of nondegenerate 2nd order Helmholtz superintegrable systems and how they are induced by B\^ocher contractions, 4) introduce   $so(4,\C) \to e(3,\C)$  contractions of Laplace systems and show how they produce conformally 2nd order superintegrable 2D time-dependent  Schr\"odinger equations.

From Theorem \ref{theorem1} we know that the potentials of all Helmholtz superintegrable systems are completely determined by their free quadratic algebras, i.e. the symmetry algebra that remains when the parameters in the potential are set equal to 0. Thus for classification purposes it is enough to classify free abstract quadratic algebras.
In a second paper under preparation we will 1) apply the B\^ocher construction to degenerate (1-parameter) Helmholtz superintegrable systems (which admit a 1st order symmetry), 
2) give a classification of free abstract  degenerate quadratic algebras and  identify which of those correspond  free 2nd order superintegrable systems. 3)  classify  abstract contractions of degenerate quadratic  algebras and identify which of those correspond to geometric contractions of Helmholtz superintegrable systems, 4) classify free abstract  nondegenerate quadratic algebras and identify  those corresponding to free nondegenerate Helmholtz  2nd order superintegrable systems, 5) classify of abstract contractions of nondegenerate quadratic  algebras.

  We note that by taking contractions  step-by-step from a model of the  $S_9$ quadratic algebra  we can recover the Askey Scheme., \cite{KMP2014}.
However, the contraction method is more general. It applies to all special functions 
that arise from the quantum systems via separation of variables, not just polynomials of hypergeometric type, and it extends to higher dimensions.
The functions in the Askey Scheme are just those  hypergeometric polynomials that arise as the expansion coefficients relating two 
separable eigenbases  that are {\it both} of hypergeometric type. Thus, there are some 
contractions which do not fit in the Askey scheme since the physical system fails to have such a pair of separable eigenbases. In a third paper under preparation we will analyze the Laplace 2nd order conformally superintegrable systems, determine which of them is exactly solvable or quasi-exactly solvable and identify the spaces of polynomials that arise. Again, multiple Helmholtz superintegrable systems will correspond to a single Laplace system. This will enable us to apply our results to characterize polynomial eigenfunctions not of Askey type and their limits.

\begin{acknowledgements}
This work was partially supported by a grant from the Simons Foundation (\# 208754 to Willard Miller, Jr).
\end{acknowledgements}

\bibliographystyle{actapoly}

\begin{thebibliography}{99}
\bibitem{EVAN}
Evans N.W.,
Super-Integrability of the Winternitz System;
{\it Phys. Lett.}\  V.A 147,  483--486, (1990), http://dx.doi.org/10.1016/0375-9601(90)90611-Q.
\bibitem{TTW2001} Tempesta P., Turbiner A. and Winternitz P.,   Exact solvability of superintegrable systems,
          {\it J. Math. Phys.}, {\bf 42}, 4248--4257 (2001), http://dx.doi.org/10.1063/1.1386927.      
\bibitem{SCQS} Superintegrability in Classical 
and Quantum Systems, 
Tempesta P.,  Winternitz P.,  Miller W.,  Pogosyan G., editors, AMS,
vol. 37, 2005,  ISBN-10: 0-8218-3329-4, ISBN-13: 978-0-8218-3329-2 
\bibitem{FORDY} Fordy A.~P.,   Quantum Super-Integrable Systems as
  Exactly Solvable Models ,  SIGMA  {\bf 3}  025,  (2007), http://dx.doi.org/10.3842/SIGMA.2007.025
 \bibitem{MPW2013}  Miller W. Jr.,   Post S. and  Winternitz P..  Classical and Quantum Superintegrability with Applications , 
  J. Phys. A: Math. Theor., {\bf 46}, 423001,  (2013)
\bibitem{KKM20041}
Kalnins E.\ G., Kress J\ .M, and  Miller W.\ Jr., 
Second  order superintegrable systems in conformally
flat spaces.  I: 2D classical structure theory. {\it J. Math. Phys.},
{\bf 46}, 053509, ( 2005); 
  II: The classical 2D St\"ackel transform. {\it J. Math. Phys.},
 {\bf 46}, 053510, (2005);  III. 3D classical structure theory, J. Math. Phys., {\bf 46}, 103507, (2005), IV. The classical 3D St\"ackel transform and 3D classification theory,, J. Math. Phys., {\bf 47}, 043514, (2006) ;  V: 2D and 3D quantum systems. {\it
  J. Math. Phys.},  {\bf 47}, 09350, (2006);
Nondegenerate 2D complex Euclidean superintegrable systems and algebraic varieties, {\it J. Phys. A: Math. Theor.},  {\bf  40},  3399-3411, (2007), 
http://dx.doi.org/10.1088/1751-8113/40/13/008.
\bibitem{DASK2007}
Daskaloyannis C.\ and Tanoudis Y.,         Quantum   superintegrable
systems with quadratic integrals 
on a two dimensional manifold. {\it J. Math Phys.}, {\bf 48}, 072108 (2007).     
\bibitem{Koenigs}
Koenigs, G., Sur les g\'eod\'esiques a int\'egrales quadratiques. A note
appearing in ``Lecons sur la th\'eorie g\'en\'erale des
surfaces''. G. Darboux. Vol 4, 368-404, 1896, {\it Chelsea Publishing} 1972. 
\bibitem{KKMW}
Kalnins E.\ G., Kress J.\ M., Miller W.\ Jr. and Winternitz P.,
{\it Superintegrable systems in Darboux spaces.}
{\it J.~Math.~Phys.},\  V.44, 5811--5848,  (2003), http://dx.doi.org/10.1063/1.1619580.
%

\bibitem{Zhedanov1992a} Granovskii Ya.\ I., Zhedanov A.\ S., and Lutsenko  I.\ M., Quadratic algebras and dynamics in curved spaces. I. Oscillator,
{\it Theoret. and Math. Phys.},  1992, {\bf  91}, 474-480, http://dx.doi.org/10.1007/BF01018846;   Quadratic algebras and dynamics in curved spaces. II. The
Kepler problem, {\it Theoret. and Math. Phys.}, {\bf  91}, 604-612, (1992), http://dx.doi.org/10.1007/BF01017335.
\bibitem{BDK}
Bonatos D., Daskaloyannis C. and Kokkotas K., 
Deformed
Oscillator Algebras for Two-Dimensional Quantum Superintegrable
Systems;
{\it Phys. Rev.},  V.A 50,  3700--3709, (1994), http://dx.doi.org/10.1103/PhysRevA.50.3700
 \bibitem{VILE}
Letourneau  P. and Vinet L.,
Superintegrable systems:
Polynomial Algebras and Quasi-Exactly Solvable
Hamiltonians.
{\it Ann. Phys.},  V.243, 144--168, (1995), http://dx.doi.org/10.1006/aphy.1995.1094
\bibitem{KKMP}
        Kalnins E. G., Kress J. M.,Miller W.  Jr. and Pogosyan G. S.,
         Completeness of superintegrability in two-dimensional constant
        curvature spaces.
        {\it J.~Phys.~ A:~Math~Gen. \bf 34}, 4705--4720 (2001), http://dx.doi.org/10.1088/0305-4470/34/22/311
\bibitem{Kress2007} Kress J.\ M., Equivalence of superintegrable systems in
  two dimensions. {\it Phys. Atomic Nuclei}, {\bf 70}, 560-566, (2007), http://dx.doi.org/10.1088/0305-4470/34/22/311
\bibitem{KM2014}Kalnins E. G.  and  Miller W. Jr.,  Quadratic algebra contractions and 2nd order superintegrable systems, 
{\it Anal. Appl.} {\bf 12}, 583-612,  (2014), http://dx.doi.org/10.1142/S0219530514500377
\bibitem{KMP2007a}   Kalnins E. G., Miller W.\ Jr.  and Post S.,
         Wilson polynomials and the generic superintegrable system on the 2-sphere,
         {\it J. Phys. A: Math. Theor. \bf 40},  11525-11538 (2007), http://dx.doi.org/10.1088/1751-8113/40/38/005
\bibitem{KMP2008} Kalnins E. G., Miller W.\ Jr.  and Post S., Models for quadratic algebras associated with second order 
superintegrable systems, {\it SIGMA} 4, 008, 21 pages; arXiv:0801.2848,(2008),  http://dx.doi.org/10.3842/SIGMA.2008.008
\bibitem{KMP2011}         Kalnins E. G., Miller W.\ Jr.  and Post S.,
       Two-variable Wilson polynomials and the generic superintegrable system on the 3-sphere, http://www.emis.de/journals/SIGMA/2011/051/ [math-ph], 
         {\it SIGMA},{ \bf 7}, 051 (2011) 26 pages,  http://dx.doi.org/10.3842/SIGMA.2011.051
\bibitem{P2011} Post, S., Models of quadratic algebras generated by superintegrable systems in 2D.{\it  SIGMA} 7 (2011),
036, 20 pages arXiv:1104.0734, http://dx.doi.org/10.3842/SIGMA.2011.036
\bibitem{LM2014} Li Q, and Miller W.  Jr, Wilson polynomials/functions and intertwining operators for the generic quantum 
superintegrable system on the 2-sphere, 2015 {\it J. Phys.: Conf. Ser.} 597 012059 (http://iopscience.iop.org/1742-6596/597/1/012059) 
\bibitem{DLMF} Digital Library of Mathematical Functions
(http://dlmf.nist.gov).
\bibitem{Wigner} In\"on\"u E.\ and Wigner E.\ P., On the contraction of groups and their representations. 
{\it Proc. Nat. Acad. Sci.} (US), {\bf 39}, 510-524, (1953), http://dx.doi.org/10.1073/pnas.39.6.510
\bibitem{WW}   Weimar-Woods E., The three-dimensional real Lie algebras and their contractions, {\it J.  Math. Phys.}, 
{\bf 32}, 2028-2033 (1991), http://dx.doi.org/10.1063/1.529222
\bibitem{NP} Nesterenko M. and 
Popovych R.,  Contractions of low-dimensional Lie algebras, {\it  J. Math. Phys.}, {\bf  47} 123515,  (2006). 
\bibitem{KMP2014} Kalnins E. G., Miller W. Jr and   Post S., Contractions of 2D 2nd order quantum superintegrable systems and the Askey 
scheme for hypergeometric orthogonal polynomials
{\em   SIGMA},  {\bf 9} 057, 28 pages, (2013),  http://dx.doi.org/10.3842/SIGMA.2013.057
\bibitem{Laplace2011} Kalnins E. G., Kress J.M, Miller W. Jr and Post, S., Laplace-type  equations as conformal
superintegrable systems,  Advances in Applied Mathematics 46 (2011) 396416.   
\bibitem{CapelKress} Capel J.J. and  Kress J.M., Invariant Classification of Second-order Conformally Flat Superintegrable Systems,
J. Phys. A: Math. Theor. 47 (2014), 495202.
\bibitem{Bocher} B\^ocher M., Ueber die  Reihenentwickelungen der Potentialtheorie, B. G. Teubner, Leipzig 1894.
\bibitem{CCM} Kalnins E,G., Miller W. Jr. and Post S., Coupling constant metamorphosis and Nth order symmetries in classical 
and quantum mechanics,   J. Phys. A: Math. Theor. 43 (2010) 035202. (20 pages) ,     doi: 10.1088/1751-8113/43/3/035202
\bibitem{Proceedings}Kalnins E.G., Miller W. Jr., and Reid G.J.,  Separation of variables for complex Riemannian spaces of 
constant curvature. I. Orthogonal separable coordinates for Snc and Enc, Proc. R. Soc. Lond. A 394 (1984),
pp. 183-206, http://dx.doi.org/10.1098/rspa.1984.0075
\bibitem{Bromwich} Bromwich T. J. A., Quadratic forms and their classification by means of invariant
factors. Cambridge tract no. 3. Cambridge University Press, 1906, reprint Hafner, New York, 1971.
\bibitem{CKP2015} Capel J.J., Kress J.M. and Post S., Invariant Classification and Limits of Maximally Superintegrable Systems in 3D,
{\it SIGMA}, {\bf 11} (2015), 038, 17 pages      arXiv:1501.06601      http://dx.doi.org/10.3842/SIGMA.2015.038


\end{thebibliography}

\end{document}